\documentclass[%
 reprint,
 amsmath,amssymb,
 aps,
]{revtex4-1}

\usepackage{graphicx}
\usepackage{dcolumn}
\usepackage{bm}

\begin{document}

\preprint{APS/123-QED}

\title{Influence of a Thermal Bath on The Transport Properties of an Open Molecular Junction}

\author{A. Eskandari-asl}
\email[Email: ]{a{\_}eskandariasl@sbu.ac.ir; amir.eskandari.asl@gmail.com}
\affiliation{Department of physics, Shahid Beheshti University, G. C. Evin, Tehran 1983963113, Iran}

\date{\today}

\begin{abstract}
In a molecular junction (MJ) which connects two electrical leads, electron-phonon coupling has significant effects on the transport properties. However, the MJ is not thermally isolated and the phonons can be coupled to another thermal bath. For strong enough couplings, the bath thermalizes phonons on the MJ so that their number would be bias independent. However, in medium and weak coupling regimes, the number of phonons created in MJ depends on the bias voltage. Obtaining the master equation (ME) for this system and comparing the results with the case where we have no such thermal bath, we show that if the bath temperature is greater than the leads, at low bias voltages (where in the absence of the thermal bath the probability of phonon excitation is low), the thermal bath heats up our MJ and decreases electronic current. On the other hand, at high bias voltages the bath cools down MJ and increases the current. However, if the bath temperature is less than the leads, it always increases the current and the heat flows from the junction to the leads. 
\end{abstract}

\maketitle

\section{Introduction}
Recent advances in nanotechnology paves the way into molecular electronics, with MJ as an important building block\cite{cuniberti,cuevas2017}. MJs which are molecular sized quantum dots (QDs) that connect electrical leads, have been the subject of many experimental and theoretical studies \cite{terrones2002,zhitenev2002,nitzan2003,galperin2006,tao2006,galperin2008,hartle2011}. 

One main difference between large QDs and MJs is that in the latter there exist a noticeable electron-phonon coupling. This coupling substantially affects transport properties of the system and may result in phenomena such as Frank-Condon blockade, negative differential resistance, etc\cite{koch2005,koch2006,galperin2005}.    

One theoretical approach is to trace out the leads degrees of freedom and obtain a ME which describes the dynamics of MJ (electrons and phonons)\cite{hartle2011,schaller2013,sowa2017,kosov2017}. In this approach, it is usually assumed that the phonons, are either completely thermalized, or just coupled to the electrons on MJ and not affected by any thermal environment other than the leads. However, regimes between these two extremes are also possible. For example, there can be a weak coupling to another thermal phonon bath (such as phonons of a substrate).

In this work we consider the coupling of MJ phonons to another thermal bath and derive the corresponding ME. Comparing the results with those obtained without such a bath, we show that depending on the bias voltage, there can be heat flow from the MJ to bath or vice versa. Moreover, when the bath temperature is higher than the leads, at low bias voltages by injecting phonons to the MJ, the bath strengthens electron blockade and reduces the electrical current. On the other hand, at high bias voltages, where many phonons would be excited by electron transport, the bath sucks in phonons from MJ and increases the current. However, in the case where the bath is cooler than the leads, the coupling increases the electrical current and the heat always flows from the MJ to the bath.

The paper is organized as follow. In Sec.\ref{mm}, we introduce the system Hamiltonian and derive the corresponding MEs. Moreover, formulas for computing electrical current and heat flow are given. In Sec.\ref{nr}, we present our numerical results and discussions, and finally, Sec.\ref{con} concludes our work.

\section{Model and Method} \label{mm}
Our model consists of a single level MJ which connects two leads. Moreover, electrons on the MJ are coupled to a single frequency phonon mode. This phonon mode is in contact to another thermal phonon bath. The Hamiltonian of this system is
\begin{eqnarray}
&&\hat{H}=\hat{H}_{m}+\hat{H}_{leads}+\hat{H}_{tun}+\hat{H}_{bath}+\hat{H}_{m-bath},
\label{hamt}
\end{eqnarray}
\begin{eqnarray}
&&\hat{H}_{m}=\epsilon_{d0} \hat{n}_{d}+\Omega \hat{b}^{\dag}\hat{b}+\lambda \Omega \hat{n}_{d} \left(\hat{b}+ \hat{b}^{\dag} \right),
\label{hm}
\end{eqnarray}
\begin{eqnarray}
&&\hat{H}_{leads}=\sum_{k,\alpha \in \left\lbrace R,L\right\rbrace } \epsilon_{k,\alpha} \hat{a}^{\dag}_{k\alpha} \hat{a}_{k\alpha} ,
\label{hleads}
\end{eqnarray}
\begin{eqnarray}
&&\hat{H}_{tun}=\sum_{k,\alpha \in \left\lbrace R,L\right\rbrace } V_{k,\alpha} \hat{c}^{\dag} \hat{a}_{k\alpha}+ h.c. ,
\label{htun}
\end{eqnarray}
\begin{eqnarray}
&&\hat{H}_{bath}=\sum_{\nu} \Omega_{\nu} \hat{b}^{\dag}_{\nu} \hat{b}_{\nu},
\label{hbath}
\end{eqnarray}  
and
\begin{eqnarray}
&&\hat{H}_{m-bath}=\sum_{\nu} \gamma_{\nu} (\hat{b}^{\dag}+\hat{b})(\hat{b}^{\dag}_{\nu}+\hat{b}_{\nu}),
\label{hm-b}
\end{eqnarray}  

where $ \hat{c} $ ($ \hat{c}^{\dag} $) is the annihilation (creation) operator of electrons on MJ, $ \hat{n}_{d}=\hat{c}^{\dag}\hat{c} $ is the number operator and $ \epsilon_{d0} $ is the onsite energy of electrons on the MJ. $ \hat{b} $($ \hat{b}^{\dag} $) is the annihilation (creation) operator of phonons on MJ, $ \Omega $ is the phonon frequency and $ \lambda $ determines electron-phonon coupling. Moreover, $ \hat{a}_{k\alpha} $ ($ \hat{a}^{\dag}_{k\alpha} $) annihilates (creates) an electron in the state $ k $ of the lead $ \alpha $ ($ \alpha=R,L $), and  $ V_{k,\alpha} $ determines the electron hopping between MJ and the leads. $ \hat{b}_{\nu} $ ($ \hat{b}^{\dag}_{\nu} $) is the annihilation (creation) operator of mode $ \nu $ of the thermal bath, $ \Omega_{\nu} $ is the energy of this mode, and $ \gamma_{\nu} $ determines the coupling strength of this mode with MJ phonons.

Performing the Lang-Firsov transformation as $ e^{\hat{S}} \hat{H} e^{-\hat{S}} $ (where $ \hat{S}\equiv \lambda \hat{n}_{d} \left( \hat{b}^{\dag}-\hat{b} \right) $ ), renormalizing the onsite energy to $ \epsilon_{d}=\epsilon_{d0} - \lambda^{2} \Omega $ and following the standard steps for deriving a Markovian ME in the limit of weak lead to MJ coupling, would result in the dynamics of the density matrix (DM) of the system (where by system we mean MJ electrons and phonons). It is straightforward to show that the rate of change of DM ($ \frac{d\rho}{dt} $) at any time is diagonal, provided that DM is diagonal at that time. Therefore, if the initial DM is diagonal, by mathematical induction we conclude that the DM will remain diagonal at all times. In such cases where all the off-diagonal terms vanish, the ME can be stated as the rate of change of the diagonal elements of DM as follow 
\begin{eqnarray}
&&\frac{d}{dt} P_{0m}=\sum_{m^{\prime},\alpha } \Gamma_{\alpha} \left( \left[ 1-f_{\alpha}\left( \Omega \left( m^{\prime}-m\right)  +\epsilon_{d}\right)\right] \vert \hat{X}_{mm^{\prime}} \vert^{2} P_{1,m^{\prime}}\right.\quad\nonumber\\
&&\left.-f_{\alpha}\left( \Omega \left( m^{\prime}-m\right)  +\epsilon_{d}\right) \vert \hat{X}_{mm^{\prime}} \vert^{2} P_{0m} \right)+\hat{\mathcal{L}}_{b} (P_{0m}),
\label{dp0}
\end{eqnarray}  

\begin{eqnarray}
&&\frac{d}{dt} P_{1m}=\sum_{m^{\prime},\alpha } \Gamma_{\alpha} \left( f_{\alpha}\left( \Omega \left(m- m^{\prime}\right)  +\epsilon_{d}\right) \vert \hat{X}_{m^{\prime}m} \vert^{2} P_{0m^{\prime}}\right.\nonumber\\
&&\left.-\left[ 1-f_{\alpha}\left( \Omega \left(m- m^{\prime}\right)  +\epsilon_{d}\right)\right] \vert \hat{X}_{m^{\prime}m} \vert^{2} P_{1,m}\right)+\hat{\mathcal{L}}_{b} (P_{1m}), \qquad
\label{dp1}
\end{eqnarray}  
where $ P_{im} $ ($ i=0,1 $) represent diagonal elements of DM and determine the probability of having $ i $ electrons in MJ and $ m $ phonons being excited. Moreover, $ \hat{X}\equiv\exp [\lambda(\hat{b}-\hat{b}^{\dag})] $, and $ f_{\alpha}(\omega)=\frac{1}{e^{\beta_{\alpha}(\omega-\mu_{\alpha})}+1} $ is the Fermi distribution of lead $ \alpha $, in which $ \mu_{\alpha} $ is chemical potential of the lead and $ \beta_{\alpha} $ is its inverse temperature. $ \Gamma_{\alpha} $ determines the tunneling rate of electrons between MJ and lead $ \alpha $, which is defined to be $ \Gamma_{\alpha}(\omega)=\sum_{k} 2 \pi \vert V_{k\alpha} \vert^{2} \delta(\epsilon_{k\alpha}-\omega) $. In wide band limit(WBL), we take $ \Gamma_{\alpha} $ to be independent of $ \omega $. $ \hat{\mathcal{L}}_{b} (P_{im}) $ is
\begin{eqnarray}
\hat{\mathcal{L}}_{b} (P_{im})&=& \Gamma_{p} \left[ 1+N_{bath}\left( \Omega \right) \right] \left[\left( m+1\right)  P_{i,m+1}-m P_{im}\right] +\nonumber \\
&&\Gamma_{p} N_{bath}\left( \Omega \right) \left[m P_{i,m-1}-\left( m+1\right)  P_{i,m}\right],
\label{lb}
\end{eqnarray}
in which $ N_{bath}\left( \Omega \right)=\frac{1}{e^{\beta_{b}\Omega}-1} $ is the number of phonons with frequency $ \Omega $ in the thermal bath (given by Bose-Einstein distribution function), in which $ \beta_{b}=\frac{1}{k_{B} T_{b}} $ is the inverse temperature of the phonon bath. Moreover, $ \Gamma_{p}=\sum_{\nu} 2 \pi \gamma_{\nu}^{2} \delta(\Omega-\Omega_{\nu}) $, determines the thermalization rate of MJ phonons.

It is noticeable that Eqs.\ref{dp0} and \ref{dp1} reduce to the ME given by Kosov\cite{kosov2017}, if we set $ \Gamma_{p}=0 $. On the other hand, if we remove the leads, these equations would determine the dynamic of a phonon system coupled to a thermal bath \cite{breuer}.   

The number of electrons in MJ is $ N_{e}=\sum_{m} P_{1m} $. The electrical currents from the leads to the MJ determine the rate of change of electron population, i.e., $ dN_{e}/dt=\sum_{\alpha} I_{\alpha} $. Comparison with Eq.\ref{dp1}, the relation for current from lead $ \alpha $ to the MJ is obtained as
\begin{eqnarray}
I_{\alpha}&=&\Gamma_{\alpha}\sum_{mm^{\prime}} \left(  -\left[ 1-f_{\alpha}\left( \Omega \left( m^{\prime}-m\right)  +\epsilon_{d}\right)\right] \vert \hat{X}_{mm^{\prime}} \vert^{2} P_{1,m^{\prime}}+ \right.\nonumber\\
&&\left.f_{\alpha}\left( \Omega \left(m- m^{\prime}\right)  +\epsilon_{d}\right) \vert \hat{X}_{m^{\prime}m} \vert^{2} P_{0m^{\prime}}\right) ,
\label{ecurr}
\end{eqnarray}
and the total current passing through the MJ is $ I=(I_{L}-I_{R})/2 $. 

The number of phonons in MJ is $ N_{ph}=\sum_{m} m \left( P_{0m}+P_{1m} \right)  $. The heating rate of MJ is $ q=\Omega ~ dN_{ph}/dt $. This $ q $ is sum of two terms $ q_{0} $ and $ q_{1} $, where $ q_{i}=\sum_{m} \Omega ~ m ~ dP_{i}/dt $. Each of these heat transfer/generation rates can be decomposed to terms that come from electronic current and thermal bath coupling. For the first part we use the index $ I $, while the second part is shown by the index $ b $. Comparison with Eqs.\ref{dp0} and \ref{dp1} suggests that we have the following relations for heat transfer rates:
\begin{eqnarray}
q_{0I}&=&\sum_{\alpha ,m,m^{\prime}} \Omega m \Gamma_{\alpha} \left( \left[ 1-f_{\alpha}\left( \Omega \left( m^{\prime}-m\right)  +\epsilon_{d}\right)\right] \vert \hat{X}_{mm^{\prime}} \vert^{2} P_{1,m^{\prime}}\right.\nonumber\\
&&\left.-f_{\alpha}\left( \Omega \left( m^{\prime}-m\right)  +\epsilon_{d}\right) \vert \hat{X}_{mm^{\prime}} \vert^{2} P_{0m} \right),
\label{q0I} 
\end{eqnarray}
\begin{eqnarray}
&&q_{1I}=\sum_{\alpha ,m,m^{\prime}} \Omega m \Gamma_{\alpha} \left( f_{\alpha}\left( \Omega \left(m- m^{\prime}\right)  +\epsilon_{d}\right) \vert \hat{X}_{m^{\prime}m} \vert^{2} P_{0m^{\prime}}\right.\nonumber\\
&&\left.-\left[ 1-f_{\alpha}\left( \Omega \left(m- m^{\prime}\right)  +\epsilon_{d}\right)\right] \vert \hat{X}_{m^{\prime}m} \vert^{2} P_{1,m}\right)+\hat{\mathcal{L}}_{b} (P_{1m}), \qquad
\label{q1I} 
\end{eqnarray}
and
\begin{eqnarray}
q_{ib}&=&\sum_{,m} \Omega m \Gamma_{p} \left[ 1+N_{bath}\left( \Omega \right) \right] \left[\left( m+1\right)  P_{i,m+1}-m P_{im}\right] +\nonumber \\
&&\Gamma_{p} N_{bath}\left( \Omega \right) \left[m P_{i,m-1}-\left( m+1\right)  P_{i,m}\right],\quad i=0,1
\label{qib} 
\end{eqnarray}
In steady state the probabilities don't change, consequently, $ q_{iI}=-q_{ib} $, which means that any heat that is produced (consumed) by electrical current, is transferred to (compensated by) the phonon thermal bath.

In order to better understand the role of thermal bath, it would be useful to assign an effective temperature to the MJ according to the Bose-Einstein distribution function
\begin{eqnarray}
T_{MJ} = \frac{\Omega}{k_{B} \ln \frac{N_{ph}+1}{N_{ph}}}.
\label{eq12}
\end{eqnarray}

With these tools, we can investigate the effect of a thermal bath on the electrical current and study the heat transfer between the bath and MJ, which is the subject of the next section.

\section{Numerical Results} \label{nr}
In this section we represent our numerical results. We work in a system of units in which $ e=\hbar=1 $. Also, the Boltzmann constant, $ k_{B} $, is taken to be 1. We set the phonon frequency to be our energy unit, i.e., $ \Omega=1 $. These automatically set our units of time and electrical current. Moreover, $ \lambda=1 $ and the bias voltage is applied symmetrically, so that $ \mu_{L}=-\mu_{R}=V/2 $, and we consider $ \epsilon_{d}=0 $. In this work, we don't consider the temperature gradient between the leads and set both leads to be at the same temperature $ T_{l} $. Finally, the tunneling rates between MJ and the leads are assumed to be $ \Gamma_{L}=\Gamma_{R}=0.01 $. 

\begin{figure}  
\includegraphics{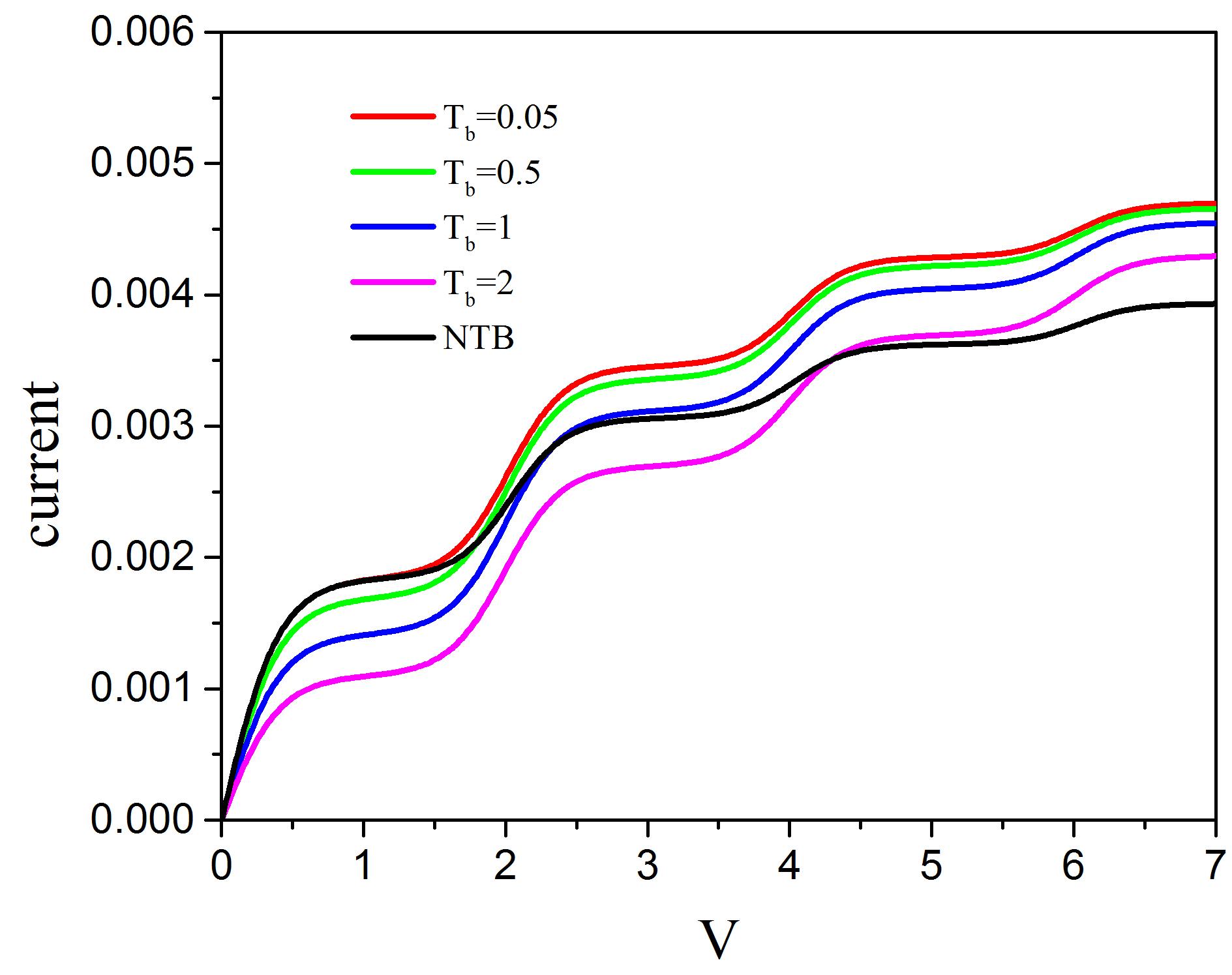}
\caption{\label{fig1} The current as a function of bias voltage, for the NTB case (solid-black curve), and the cases where a thermal bath is coupled to the MJ. The lead temperature is $ T_{l}=0.1 $ and the temperatures of thermal bath are indicated in the plot. For the cases where we have thermal bath, $ \Gamma_{p}=0.01 $. }
\end{figure}

 \begin{figure}  
\includegraphics{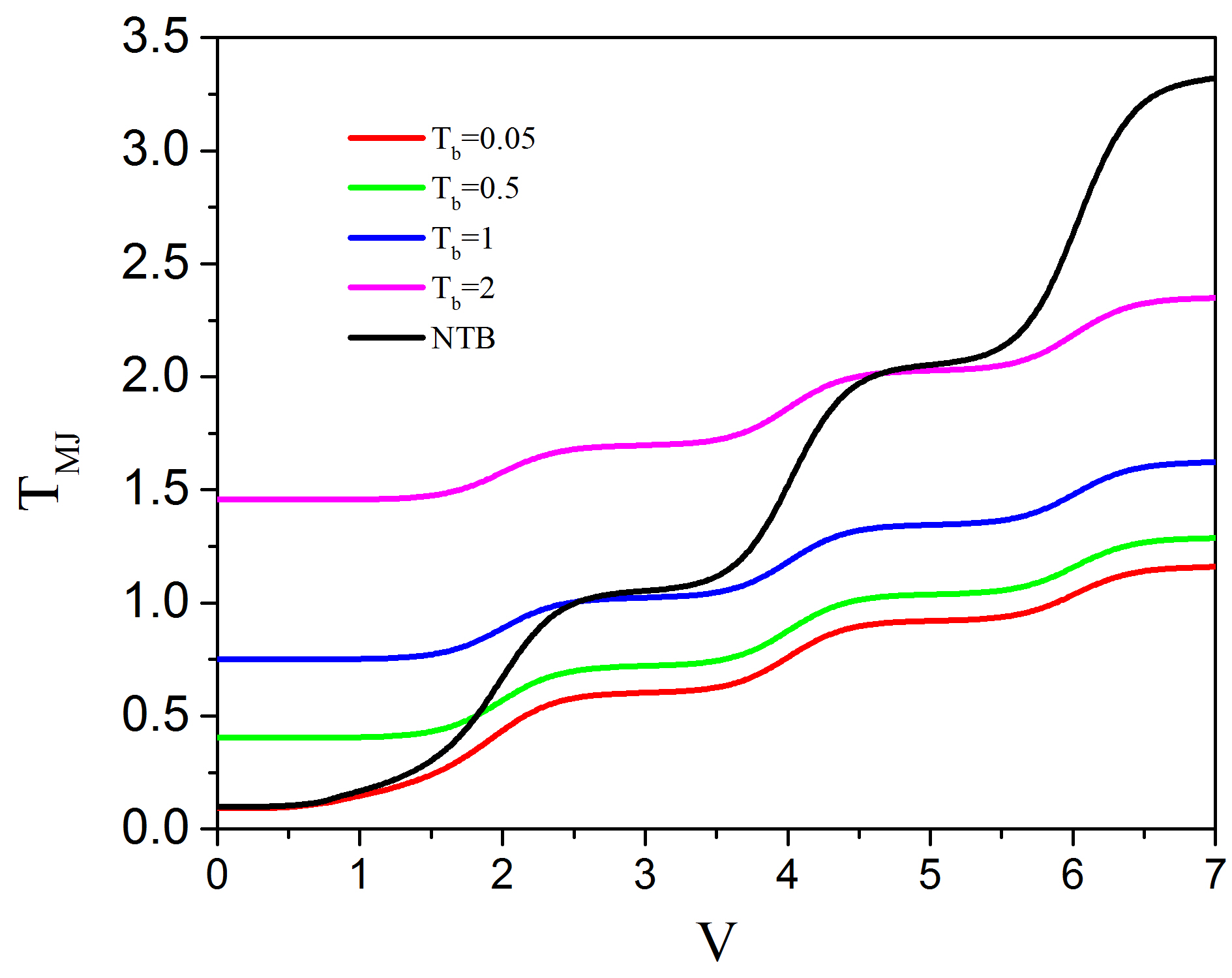}
\caption{\label{fig2} The effective temperature of the MJ, $ T_{MJ} $, as a function of bias voltage for the same situations as Fig.\ref{fig2}.}
\end{figure}

The first case we investigate is when the leads temperature is low. Fig.\ref{fig1} shows the current through MJ as a function of bias voltage, for the lead temperature $ T_{l}=0.1 $. In this figure, the solid-black curve corresponds to the case where we have no thermal bath coupled to our MJ (the NTB case). Other curves represent the situation in which the MJ is also coupled to the thermal bath with coupling strength determined by $ \Gamma_{p}=0.01 $. As it is shown in this figure, when the temperature of bath, $ T_{b} $, is greater than the lead temperature $ T_{l} $, up to some critical bias voltage (which we show by $ V_{c} $) the current is less than its value for the NTB case, i.e., the thermal bath reduces the current. However, for higher bias voltages, the current is greater than that of NTB case and the thermal bath increases the current. On the other hand, for the case where $ T_{b}<T_{l} $, the current is always greater than the NTB case. In order to understand this behavior, in Fig. \ref{fig2}, the effective temperature of MJ, $ T_{MJ} $, is depicted for similar parameters as Fig.\ref{fig1}. Electron current through MJ excites phonons and increases $ T_{MJ} $. As long as $ T_{MJ} $ in the NTB case is less than $ T_{b} $, coupling to the bath results in a heat flow from the bath to the junction, which means there would be more phonons to resist electron current (in other words, the so called Franck-Condon blockade is increased). Moreover, this heat flow increases $ T_{MJ} $ with respect to the NTB case. On the other hand, when by increasing the bias voltage and exciting more phonons in the MJ, $ T_{MJ} $ in the NTB case gets larger than $ T_{b} $, the direction of heat flow would become from the MJ to the bath, which means less phonons and consequently less resistance and lesser $ T_{MJ} $.  This discussion clarifies that $V_{c} $ is the voltage at which $ T_{MJ} $ in the NTB case equals $ T_{b} $. For the case where $ T_{b}<T_{l} $, even at zero bias  $ T_{MJ}>T_{b} $, and increasing the voltage just increases $ T_{MJ} $, therefore the heat flow is always from MJ to the bath, and the current is always greater than the NTB case. 
\begin{figure}  
\includegraphics{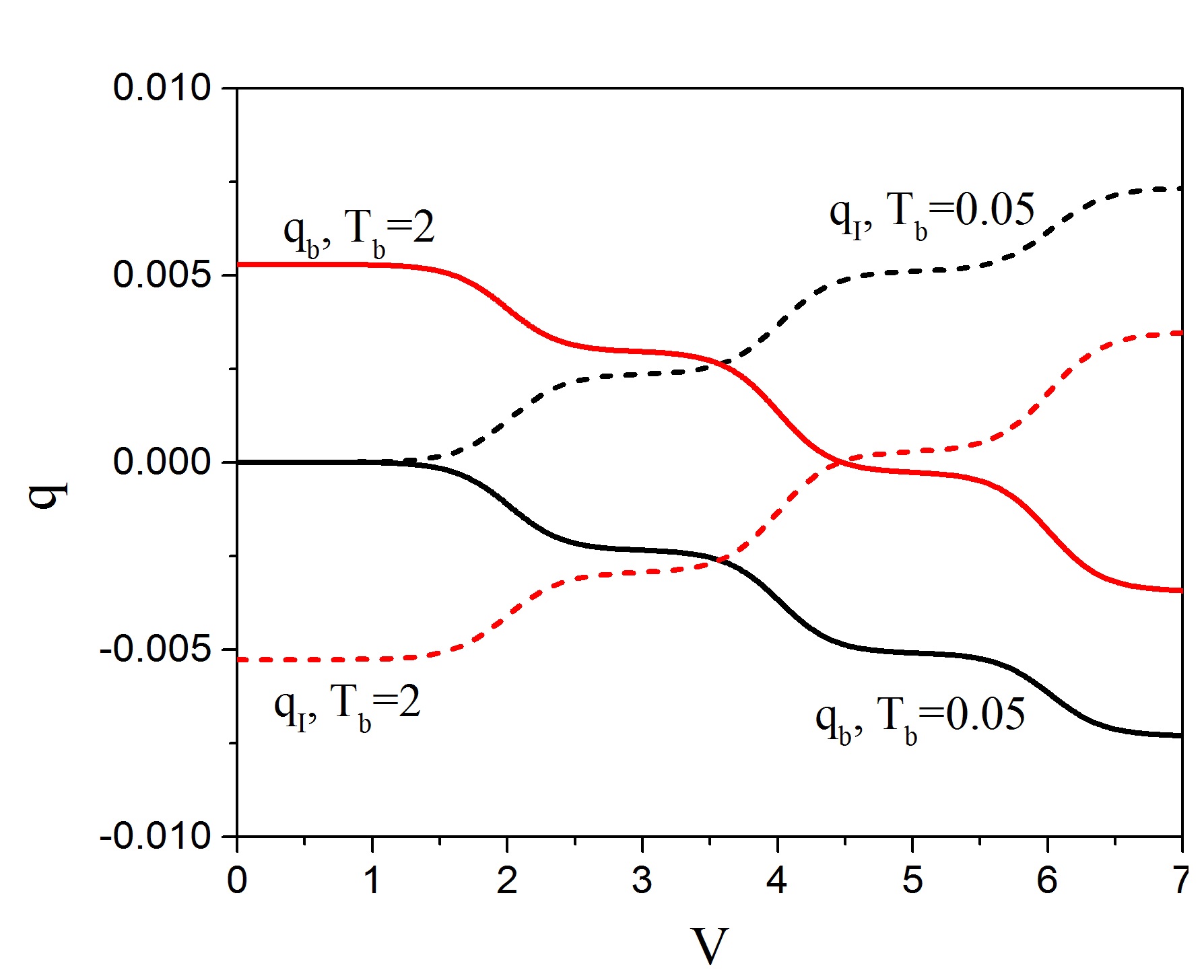}
\caption{\label{fig3} The rates of MJ heating due to current ($ q_{I} $) and thermal bath ($ q_{b} $) as functions of bias voltage for $ T_{l}=0.1 $ and $ T_{b}=0.05 $, $ 2$. Other parameters are same as Fig.\ref{fig1}.}
\end{figure}

In Fig.\ref{fig3} the rates of MJ heating due to current ($ q_{I} $) and thermal bath ($ q_{b} $) are depicted as functions of bias voltage for $ T_{l}=0.1 $ and $ T_{b}=0.05 $, $ 2$. As it is seen, the signs of heat flows are in consistence with the forgoing discussion. Moreover, for $ T_{b}=2 $ at the critical bias voltage  $ V_{c} $, both $ q_{I} $ and $ q_{b} $ vanish and there would be no heat flow between MJ and bath. For $ T_{b}=0.05 $ where the lead temperature is greater than bath, $ q_{b} $ is always negative, which means that the heat flows from the MJ to the leads. On the other hand, $ q_{I} $ is always positive in this case and electron current heats the MJ up.   

\begin{figure}  
\includegraphics{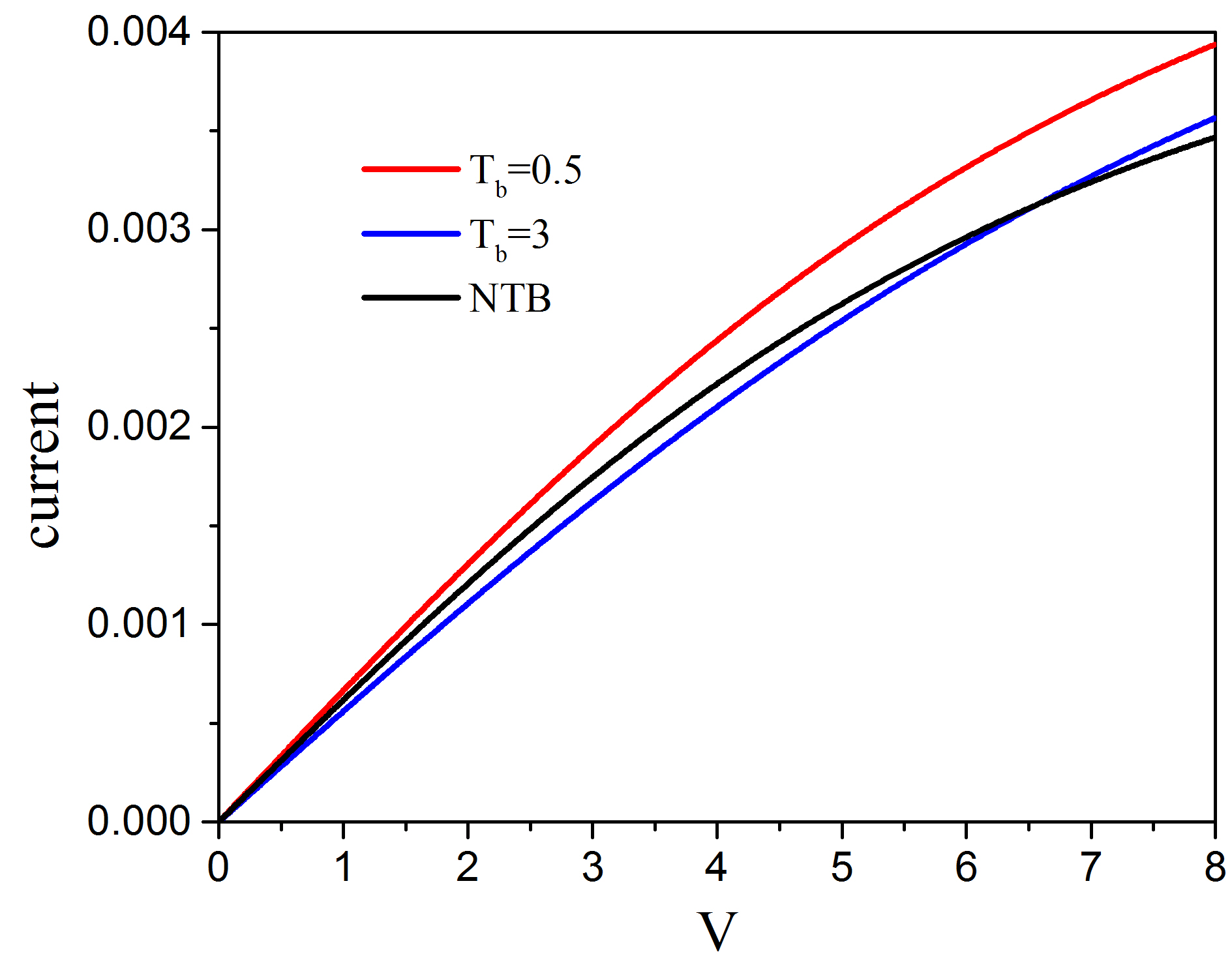}
\caption{\label{fig4} The current as a function of bias voltage, for the NTB case (solid-black curve), and the cases where a thermal bath is coupled to the MJ. The lead temperature is $ T_{l}=1.5 $ and the temperatures of thermal bath are indicated in the plot. For the cases where we have thermal bath, $ \Gamma_{p}=0.01 $.}
\end{figure}

\begin{figure}  
\includegraphics{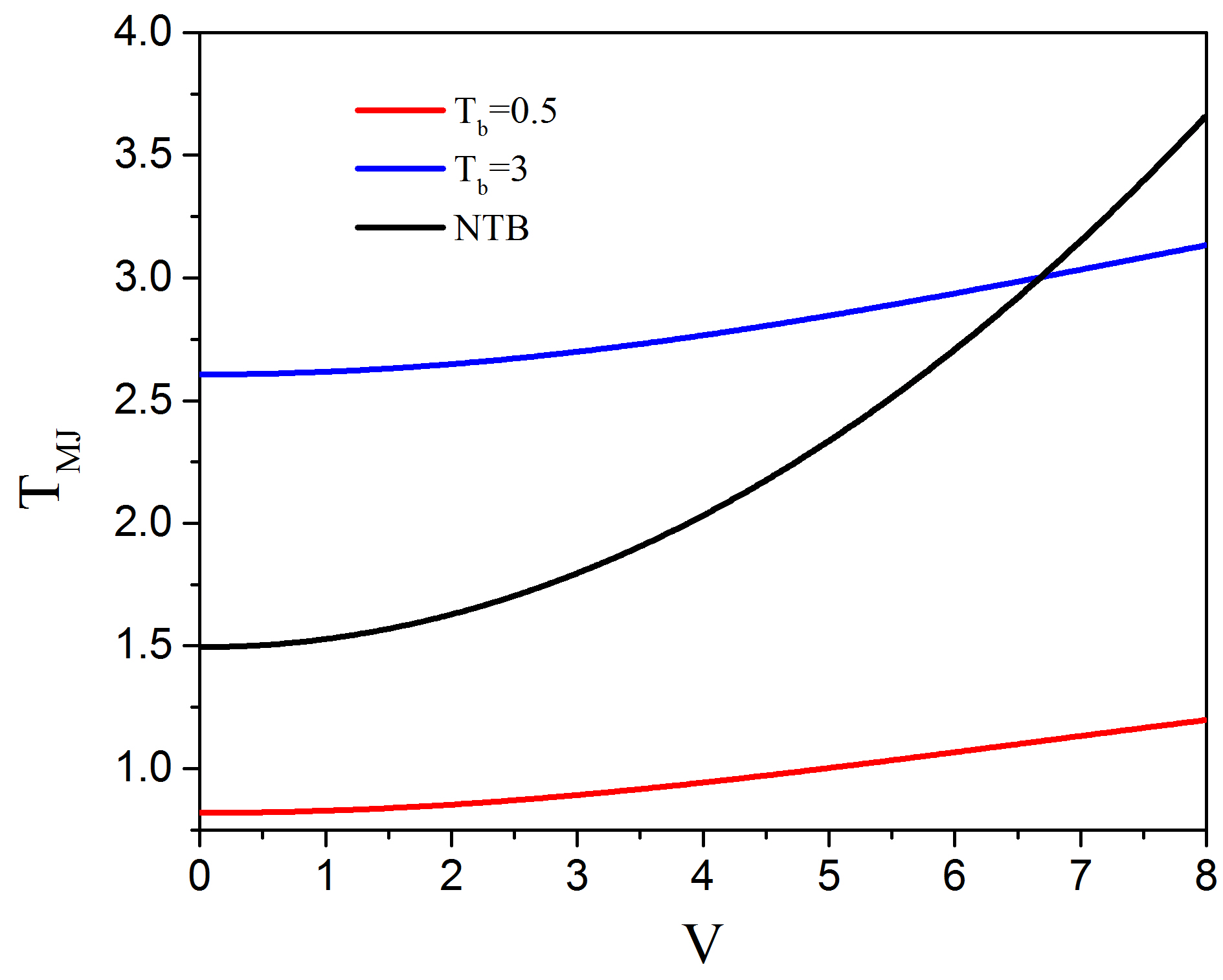}
\caption{\label{fig5} The effective temperature of the MJ, $ T_{MJ} $, as a function of bias voltage for the same situations as Fig.\ref{fig4}.}
\end{figure}

 The same behavior is seen when the leads are at high temperature, except that since the electrons get distributed around the chemical potential of leads, the step-wise behavior that is the finger print of the phonon side-bands disappears. In Fig.\ref{fig4} the current and in Fig.\ref{fig5} the effective temperature of MJ as functions of bias voltage are shown for $ T_{l}=1.5 $ and $ T_{b}=0.5 $ and 3. When $ T_{b}<T_{l} $ the bath sucks the phonons from MJ, therefore the current is increased while $ T_{MJ} $ is decreased with respect to the NTB case. On the other hand, for $ T_{b}>T_{l} $, if the bias voltage is lesser (greater) than a critical voltage $ V_{c} $, the current is decreased (increased) and $ T_{MJ} $ is increased (decreased) with respect to the NTB case.

\begin{figure}  
\includegraphics{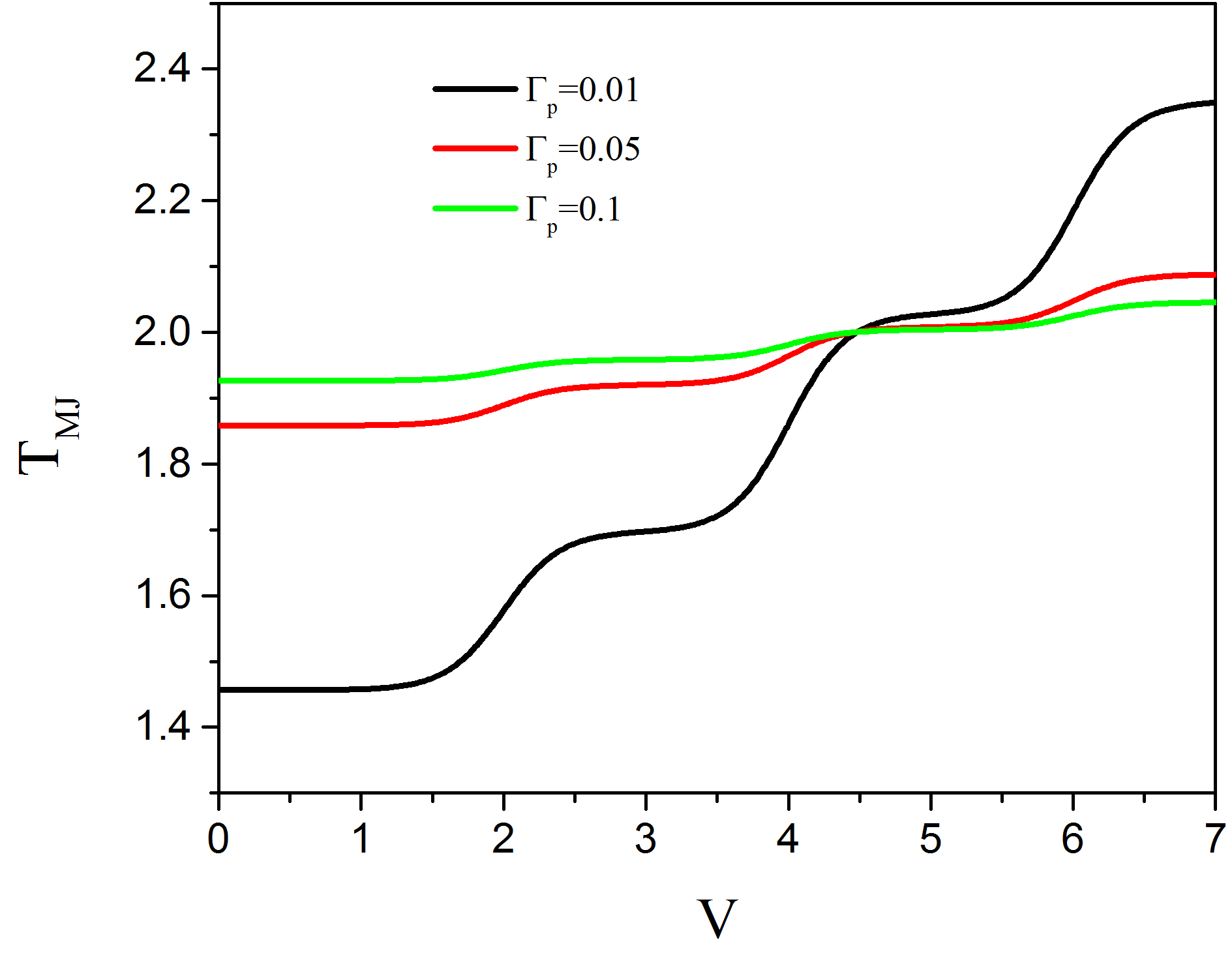}
\caption{\label{fig6} The effective temperature of the MJ, $ T_{MJ} $ as a function of bias voltage, for the cases where we have a thermal bath coupled to the MJ with different coupling strengths that are determined by $ \Gamma_{p}=0.01 $, $ \Gamma_{p}=0.05 $ and $ \Gamma_{p}=0.1 $ . The leads are at temperature $ T_{l}=0.1 $ while the bath temperature is $ T_{b}=2 $.}
\end{figure}

We expect that by increasing $ \Gamma_{p} $ (or in other words, increasing the coupling strength between the bath and MJ), the phonons on our MJ get thermalized. This means that $ T_{MJ} $ becomes less voltage dependent and stays close to $ T_{B} $. This behavior is depicted in Fig.\ref{fig6} where $ T_{MJ} $ is plotted for low temperature leads ($ T_{l}=0.1 $) and bath temperature of $ T_{b}=2 $, for three cases of $ \Gamma_{p}=0.01 $, $ \Gamma_{p}=0.05 $ and $ \Gamma_{p}=0.1 $.

\begin{figure}  
\includegraphics{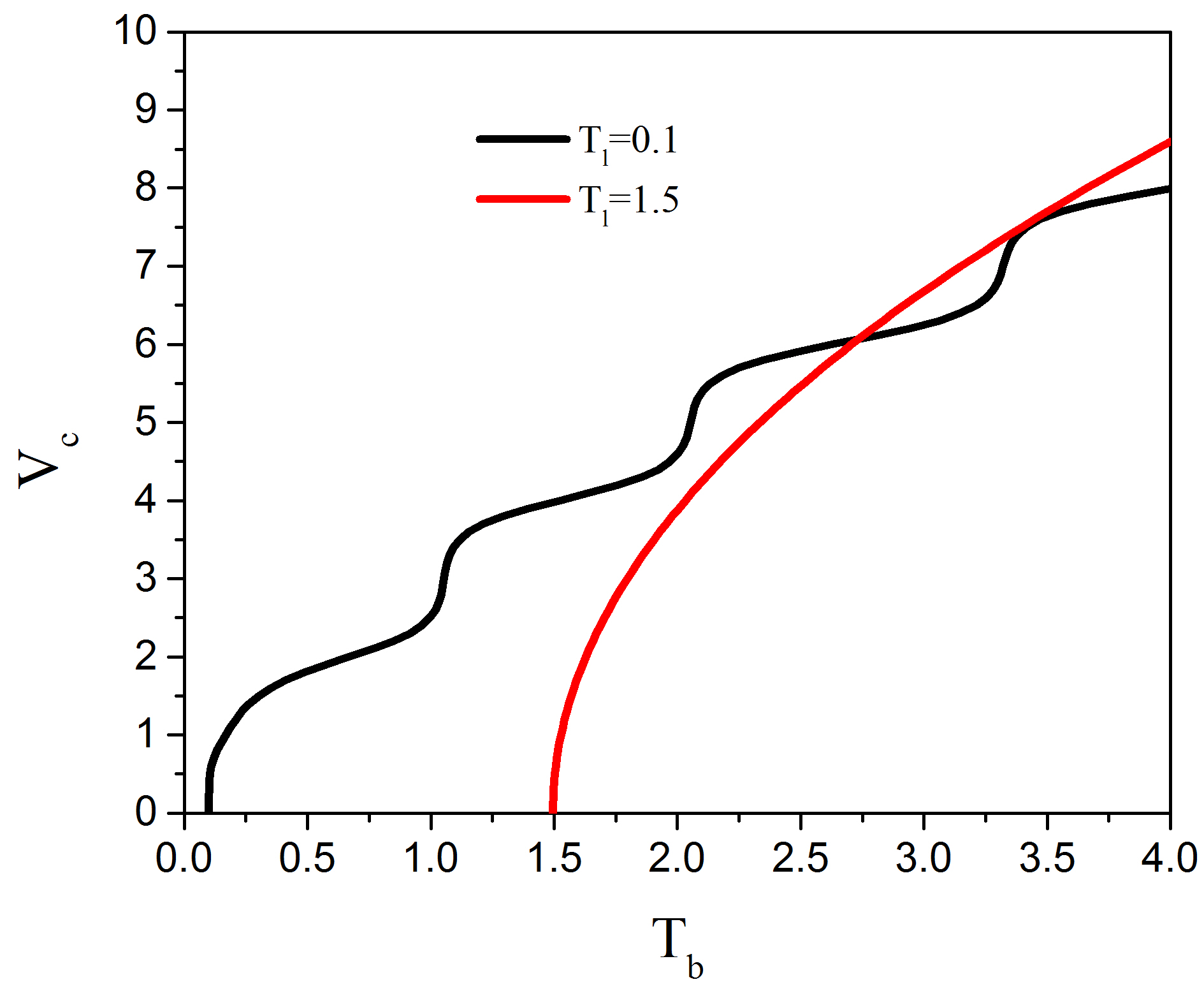}
\caption{\label{fig7} The critical voltage $ V_{c} $, as a function of bath temperature for two lead temperatures  $ T_{l}=0.1 $ and $ 1.5 $.}
\end{figure}

As we already mentioned, the critical voltage $ V_{c} $ is the voltage at which the effective temperature of MJ in the NTB case equals the bath temperature (and it is independent of $ \Gamma_{p} $). Consequently, by computing $ T_{MJ} $ as a function of bias voltage and inverting the curves, one can obtain $ V_{c} $ as a function of $ T_{b} $ for a given lead temperature. This is done in Fig.\ref{fig7}, for two values of $ T_{l}=0.1 $ and 1.5. For the bath temperatures below $ T_{l} $, there is no such critical voltage.

\section{Conclusions} \label{con}
In conclusion, we considered a current carrying MJ coupled to a single frequency phonon mode that can be either isolated from other environments or be coupled to another phononic thermal bath. For the bath temperatures greater than the leads temperature, we showed that at bias voltages lower than a critical value $ V_{c} $, the thermal bath would heat up the MJ and reduces the current. However, for higher bias voltages, the thermal bath cools down the MJ and  increases the current. This critical bias voltage is the bias voltage at which the effective temperature of MJ in the NTB case is equal to the bath temperature. On the other hand, if the bath temperature is less then the leads, there would be always a heat flow from the MJ to the bath, and the bath helps increasing the electrical current through MJ. This behavior is seen for both low and high temperature leads, with the difference that in the former, the step-wise behavior in the current-voltage characteristic which stems from the phonon side-bands is clearly seen. 

If we increase the coupling strength of the MJ phonons with the thermal bath, $ T_{MJ} $ looses its dependence on bias voltage and stays close to $ T_{b} $, or in other words, the phonons get thermalized, as we physically expect.   
\acknowledgments
We acknowledge A. T. Rezakhani for usefull discussions. 


\bibliography{apssamp}

\end{document}